# The Origins of Blast Loaded Vessels

Jonathan E Morgan

Los Alamos National Laboratory

Los Alamos, NM 87545

**Abstract**: As the Manhattan Project shifted to the theory of implosion assembly in 1944, plutonium was extremely rare and large uncertainties surrounded the function of the Gadget. For these reasons, a team within the Manhattan Project began another ambitious experiment: to confine the effects of detonating two tons of high explosives and enable the recovery of precious plutonium! No data existed on the subject, and the team faced numerous challenges as they engineered what is believed to be the world's first blast-loaded confinement vessel.

## Introduction

Project Y, a secret laboratory established by the Manhattan Project,[1] kicked off in Los Alamos, NM, in 1943 and work began immediately on a gun-assembly design known as Thin Man, a 17-foot-long plutonium gun-assembled weapon. Figure 1 shows Thin Man bomb cases.

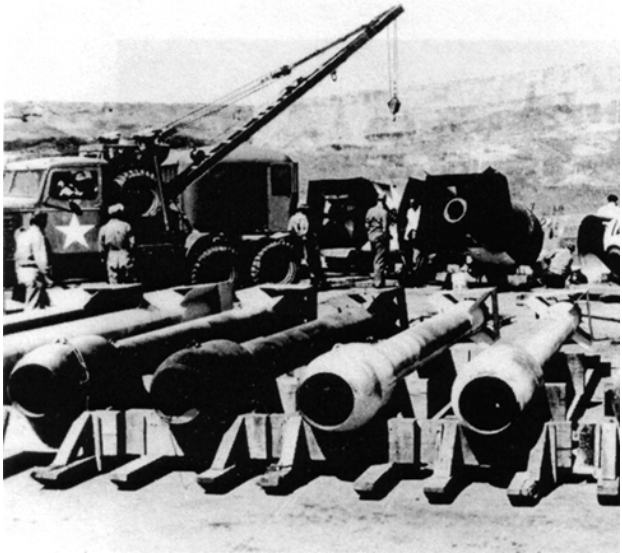

Figure 1. Thin Man bomb cases at Wendover Army Airfield in Nevada, circa early 1944. Each test body had a different center of gravity to test flight stability.

Before we dive deeper into the origins of the blast-loaded vessel, we have to rewind to 1940. In 1940, research was underway in the Radiation Laboratory at the University of California Berkeley on radioactive elements using a 60-inch cyclotron, as shown in Figure 2, to bombard uranium atoms with deuterons. Edwin M.

McMillan and Philip H. Abelson were studying these uranium fission fragments and identified element 93, neptunium. In December 1940, Glenn T. Seaborg revealed that an isotope of neptunium decayed to another transuranium element, atomic number 94, which he later named plutonium and gave the symbol Pu. By May 1941, he estimated that plutonium-239 was 50% more likely than uranium-235 to fission, based on the fission cross-section,[2,3,4] making it a better choice than uranium to create a nuclear weapon. Emilio Segrè and Glenn Seaborg successfully produced 28 μg of plutonium in the 60-inch cyclotron, but using the cyclotron to produce weapon-usable amounts of plutonium was not practical.

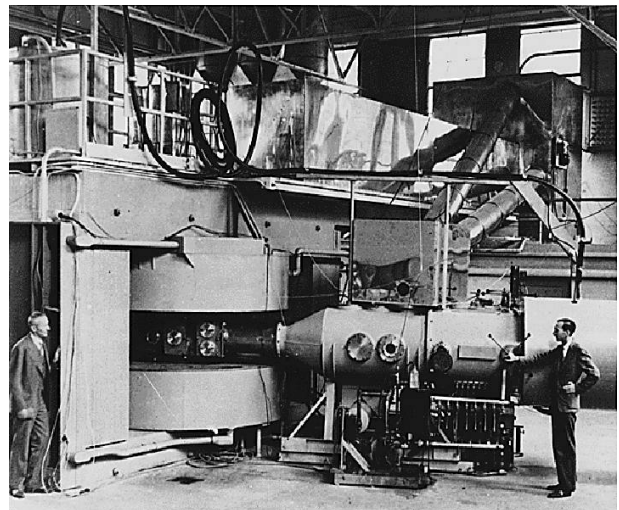

Figure 2. The 60-inch cyclotron in the University of California's Berkeley Radiation Laboratory, operational in 1939.

Glenn Seaborg isolated a weighable sample of plutonium in August 1942 at the University of Chicago Metallurgical Laboratory, using irradiated material from

---

Chicago Pile Number 1. This created a way to manufacture weapon-usable quantities of plutonium. Based on these discoveries, construction of the world's second and third nuclear reactors commenced. The X-10 graphite-moderated reactor building, shown in Figure 3, was constructed at Oak Ridge, TN, and the first sample of plutonium from the reactor was sent to Los Alamos in April 1944.

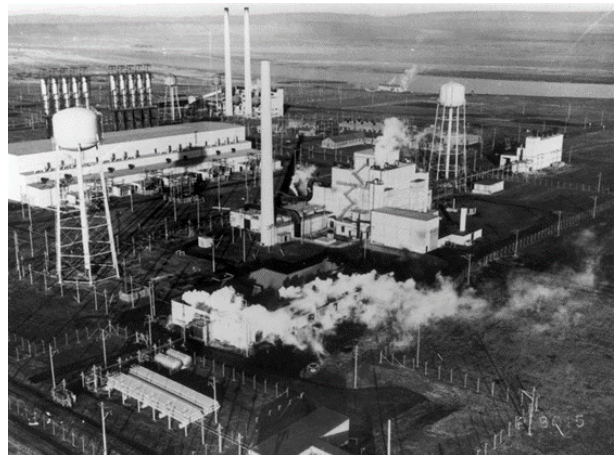

Figure 4. Hanford Reactor B, center right. The Queen Mary chemical separation facility is visible in the upper left. First critical on Sep 26, 1944.

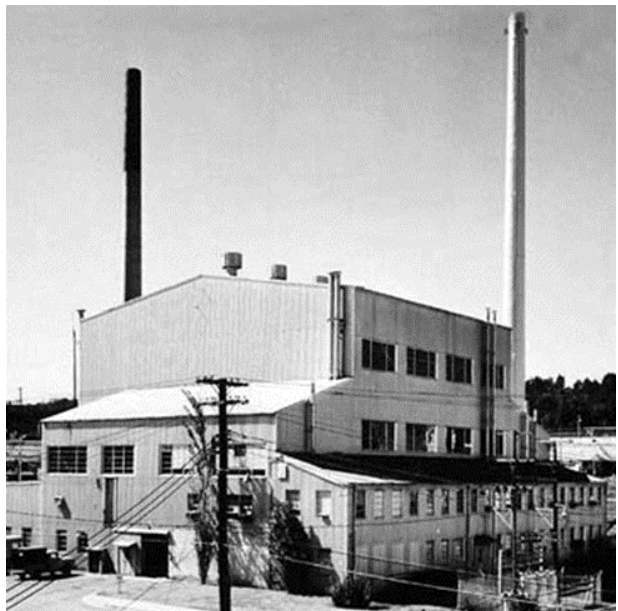

Figure 3. The X-10 graphite-moderated nuclear reactor building in Oak Ridge, TN. First critical on Nov 4, 1943.

Based on lessons learned from the X-10 pile, a full production scale reactor, Reactor B, was built in Hanford, WA, on the banks of the Columbia River, along with a huge chemical separation plant that became known as the Queen Mary. Reactor B provided its first plutonium to Los Alamos in February 1945. Reactor B is shown in Figure 4.

Analysis of the reactor-based plutonium revealed a new problem. P Division, at the time one of a handful of divisions within Project Y, under Emilio Segrè, discovered plutonium-240 mixed with plutonium-239, dramatically increasing the spontaneous fission rate of the mixture and thus eliminating the possibility that the Thin Man gun-assembly method would function as designed. This discovery forced the team at Los Alamos to shift the major focus of the project to the implosion method for plutonium, instead of the gun method. The implosion design that the team settled on became known as the Christy Gadget, after inventor Robert Christy.[5] With this revelation, Robert Oppenheimer, the director of Project Y, made the decision to restructure the entire workforce at Los Alamos to pursue parallel paths: a uranium gun assembly based on a design that became known as Little Boy and the implosion method based on the Christy Gadget. Thus, in the spring of 1944, merely one year into its existence, Los Alamos initiated a tradition—reorganizations! Ordnance (E) and Experimental Physics (P) Divisions were disbanded, and three new divisions were created. Gadget (G) Division was created to pursue weapons physics, Explosives (X) Division was created to develop implosion methods using high explosives (HE), and a new Ordnance (O) Division was created to turn the two weapon assembly methods into militarily deliverable weapons. With the extremely limited supply of plutonium, making gold look rather pedestrian, and the shift to the implosion method, concern was raised about the possibility of wasting the precious plutonium in a fizzle, or dud, so a new project within Project Y was initiated. The challenge was to capture and recover plutonium from a fizzle yielding device. The capture of the precious plutonium is the focus of this paper.

**Plutonium Capture Methods**

Based on historical research conducted on the Project Y records, four methods of plutonium capture were simultaneously considered: detonate the test device in a massive underground reinforced concrete "bomb proof," detonate it in a large sand pile, detonate it in a large "water baffle," or detonate it in a confinement vessel. Ironically, all of these methods would see some form of use to confine or contain hazardous materials from HE detonations over the next 75 years of nuclear weapons research and development.

Discussions were held with Aberdeen Proving Grounds, located near Aberdeen, MD, on the feasibility of using a bomb proof for confinement, but no further action was taken during the project.[6] The bomb-proof concept eventually became a reality with the construction of the Contained Firing Facility at LLNL Site 300, which became operational around 2000.

The sand-pile method did not proceed beyond scaled experiments, but the concept evolved to a system known as a "Gravel Gertie" for confinement of an accidental detonation in a weapon assembly bay. See Figure 5 for a Nevada Test Site shot of a mock Gravel Gertie.

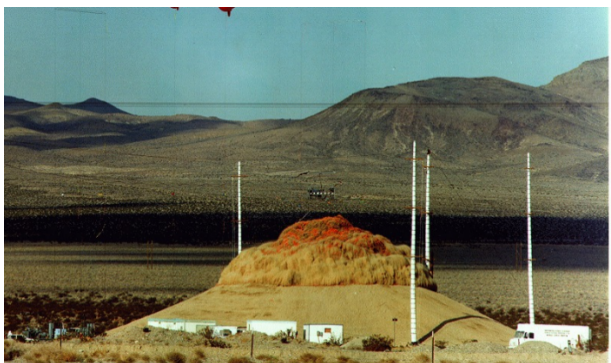

Figure 5. November 1982 Gravel Gertie shot with 423 lb HE at the Nevada Test Site[7].

For the water-capture method, the team proposed that thousands of gallons of water would encase the device over a large, cupped concrete basin. After the detonation, the debris would be collected and processed to collect the precious plutonium. This method moved from theoretical concept to scaled detonation tests. Initial tests were performed on an asphalt strip, as shown in Figure 6.

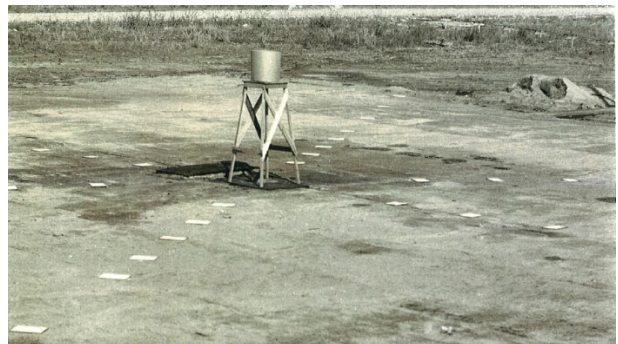

Figure 6. Initial tests of the water-capture method on an asphalt strip, circa early 1944.

After favorable results with cobalt liners in steel spheres with tuballoy[8] tampers, the water-capture method was scaled up to a larger concrete bowl at LANL's TA-6 site. Figure 7 was taken shortly after the completion of the bowl. This photo was recently discovered at the National Security Research Center (NSRC) archives and declassified for this article, along with a host of other photographs and documents. R.W. Carlson stated, "Using a water to High Explosive (HE) ratio of 50 to 1, 86% of a cobalt liner was recovered within a 30-foot-radius circle for a 1/10th-scale shot placed approximately three feet above the ground."[9]

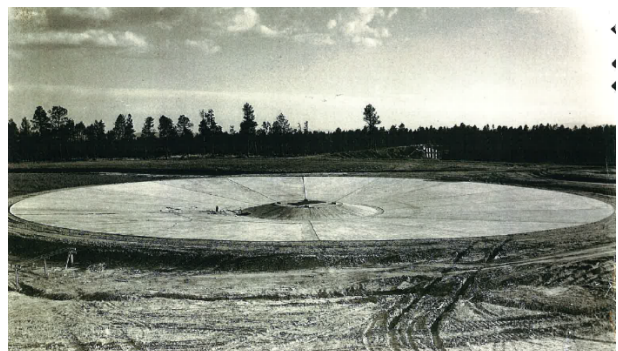

Figure 7. Concrete bowl for recovery of water shots at TA-6, circa late 1944. The bowl is now part of the Manhattan District National Park.

In the early 2000s, a similar method was employed at LANL's Dual Axis Radiographic Hydrodynamic Test (DARHT) facility at TA-15, where fire-fighting-type foam was used to reduce airborne beryllium concentrations from each shot. Figure 8 shows the complex "tent" structure used to confine the foam for a shot. Postshot cleanup work was substantial, and the process was dropped in 2007. Instead, thick-walled, high-strength, fracture-tough steel confinement vessels were chosen for use. Field experience has shown that there are significantly fewer logistical challenges using confinement vessels.

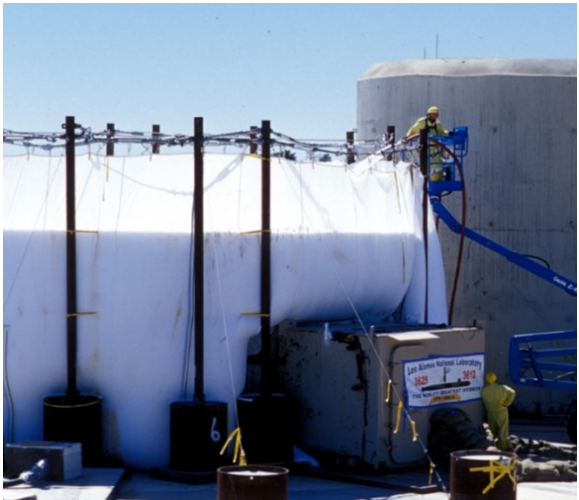

Figure 8. Filling a confinement "tent" with water-based foam prior to a hydrodynamic shot at DARHT. Foam shots were conducted from 2001-2005 before moving to vessels in 2006.

Similar to the move to confinement vessels at DARHT, the Project Y team made the decision to pursue confinement vessels simultaneously with the water recovery method.

### Confinement of an Explosion by a Steel Vessel

The team set initial requirements for the confinement of two tons of explosives. The first challenge was a complete lack of data on the subject. No blast pressure data for large explosive charges at close distances existed, so the team started by extrapolating from data gathered by the National Research Council's Passive Protection Against Bombing (PPAB) committee. The committee had collected data on bombing effects for concrete, soil, framed houses, and basements. The study ended just as the Project Y team was starting on the development path for a confinement vessel in June 1944. With extrapolations from this data, construction began on an assortment of small cast-steel spherical vessels, nicknamed Jumbinos, for dynamic testing.

These vessels ranged from 5-inch to 24-inch inside diameters (IDs) with diametrically opposed, threaded access ports constructed from 30,000 psi and 60,000 psi yield strength cast steel.[10] Detailed quantities for the smaller vessels have not been identified, but documentation is clear for an order of eight 24-inch cast-steel spheres. The order was placed with the Lebanon Steel Foundry in Pennsylvania for four at the lower yield strength and four at the high yield strength. The Project Y team objected to the foundry technique proposed and tried to convince Lebanon Steel to use an alternative method, but Lebanon Steel did not agree with the request. Metallurgical reports confirmed defects in the vessels, and no additional orders were placed with Lebanon Steel as the team considered them incompetent.[11]

At least 79 shots were conducted in the 6-to-12-inch cast-steel vessels alone.[12] The total number of shots fired may never be known. See Figure 9 for an example of a test setup, and Figure 10 for an example of a failed test shot. Radial expansion of the vessels was measured for various charge weights and types. For example, a successful vessel test was conducted with 1.75 lb of Pentolite, and another vessel failed with 2 lb of Pentolite.[13] The reason for the failure was not provided but a reasonable conclusion is casting imperfections initiated the failures.

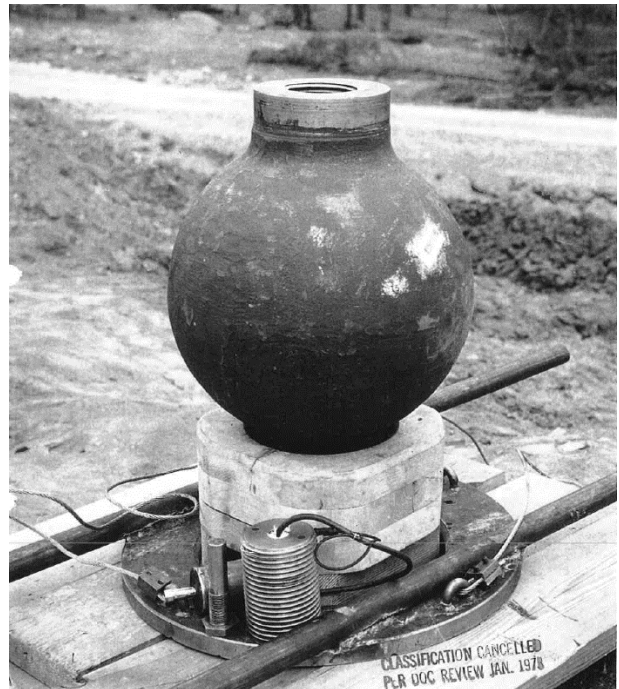

Figure 9. Cast-steel vessel test setup, early 1944. Note threaded plug with cable feedthrough.

---

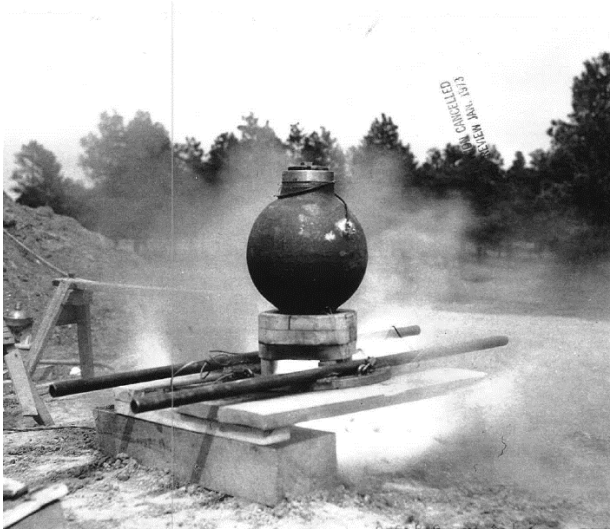

Figure 10. Cast-steel vessel failure from the bottom plug/manway, early 1944.

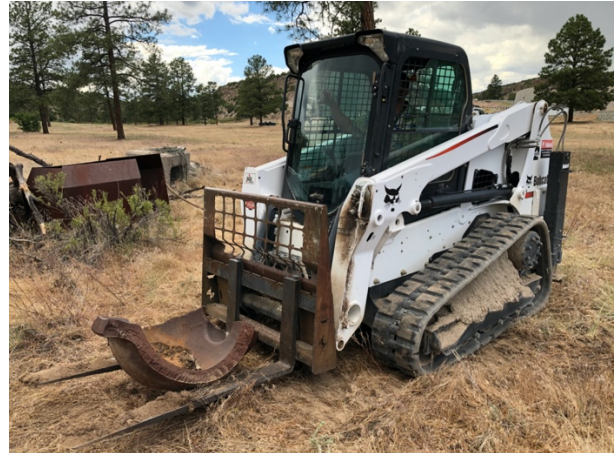

Figure 11. Recovery of a large fragment from one of the 24-inch cast-steel Jumbinos, May 29, 2020.

G.B. Kistiakowsky, H.A. Bethe, E. Teller, J. Von Neumann, and K.T. Bainbridge were struggling to predict the destructive effects of the shock wave on the containing sphere and the properties to design against (tensile strength, yield strength, ductility, and/or inertial confinement). The vessel effort reached its lowest point to date in March 1944, as the experimental pressures were much higher than expected, 1.6 times the initial extrapolation. This data challenged the notion of successfully confining the impulse of the full-scale Gadget in a steel vessel.[14] In the end, the team settled on a design with inertia and ductility as the primary criteria to absorb the impulse and sustained pressure (i.e., for a few seconds). The sustained pressure was calculated to be between 5000 and 8000 psi from 4500 lb of Composition B.[15]

As historical research began on the origins of blast-loaded vessels, remnants of old experiments were known to exist at the edge of the Lower Slobbovia firing site, at LANL's TA-36, but the age and purpose of the experiments were not known. Through this historical research project, the purpose of the experiments has been determined, and large fragments from one of the 24-inch ID cast-steel vessels have been recovered from the Lower Slobbovia field and the mesa above the firing site. Discussions are now underway on how to incorporate this important part of history into the Manhattan District National Park. The two-port geometry of these early cast-steel vessels can be seen in Figure 11.

In addition to extensive work on pressure loading, the team was busy performing a detailed study of fragments likely to occur from the Gadget inside Jumbo #1 as well as fragments of Jumbo #1 if the Gadget produced 50 to 500 ton TNT equivalent conditions.[16] Despite an acknowledgement regarding the uncertainty in their ability to predict the correct ballistic coefficient for the fragments, the team calculated a worst-case range safety scenario of 10,000 yards at Trinity. Using that analysis, structures beyond 10,000 yards were deemed safe from fragments.

## Jumbo #1

The team started the quest for a confinement vessel with a broad range of potential sizes for a full-scale cast-steel "Jumbo," henceforth referenced as Jumbo #1, that would be 13 to 18 feet in spherical diameter, have a wall thickness of up to 2 feet, and weigh from 80 to 250 tons. Loading conditions and the corresponding requirements for the vessel were still in a great deal of flux as data rolled in from scaled experiments.[17] Meanwhile, other members of the team scoured the country for potential vendors but were only able to identify three vendors in the entire United States that were equipped to manufacture a cast-steel sphere in these sizes.

The three companies were General Engineering and Foundry Company headquartered in Pittsburgh, PA with a suitable foundry located in Newcastle, IN, Bethlehem Steel Corporation, located in Bethlehem, PA, and Jones and Laughlin headquartered in Pittsburgh, PA with a suitable foundry located in Cleveland, OH. Team members met with representatives from all three companies and left the meetings with optimism about building the cast-steel vessel. The team dropped Jones and Laughlin from consideration because their representatives were vague on the exact method they would employ to fabricate the vessel. General Engineering and Foundry Company and Bethlehem Steel Corporation both recommended tapered cast hemispherical shells joined by a heavy circumferential thermite weld around the equator.

The team called the chief engineer of the Metal & Thermit Corporation, located in New York, NY, to discuss the feasibility of the process, and he stated, "A job of this magnitude would be of a pioneering nature."[18] With the initial inquiries into the cast-steel vessels complete, the engineers at the respective companies dug into the project, and they came back with serious concerns on the feasibility of meeting project specifications, such as porosity free castings in this scale and successfully completing the thermite weld.

On May 3, 1944, F.H. Hirschland of Metal & Thermit Corporation[19] and H.F. Weaver of Bethlehem Steel Company sent independent memos to Oppenheimer expressing these concerns.[20] Based on these memos and the realistic schedules communicated to complete this ambitious project, Oppenheimer cancelled the effort to cast a large containing sphere in early June 1944.[21] The vessel project reached a new low point. The team returned to the drawing board and started down a new path designed around plate steel and electrically welded joints, instead of cast steel and thermite welds. The Jumbo #1 concept never made it past the 24-inch Jumbinos.

### Elongated Jumbinos

The Jumbo #2 concept started life in the form of small 1015/1020 mild-steel cylinders with forged hemishell end caps. These confinement vessels were nicknamed "elongated Jumbinos." Fabrication of the elongated Jumbinos was done in parallel with the cast-steel Jumbinos to conserve time and provide concurrent solutions should one design fail. This turned out to be a very good idea. As the project progressed, "elongated" was dropped from the vernacular, causing the author some confusion while conducting the research on the project. The design for the elongated Jumbinos was a $1/10^{th}$-scale version of a new confinement vessel that the Project Y team identified as Jumbo #2 (as opposed to the cast spherical Jumbo #1 proposal, which was cancelled). The elongated Jumbinos were 18-inches long with an internal diameter of 12 inches and a 0.6-inch wall thickness. The elongated Jumbinos were nominally rated for 4.5 lb of HE with permanent plastic deformation. The first elongated Jumbinos had concrete reinforcing bands, as shown in Figure 12, but those did not perform as desired so the design was changed again.

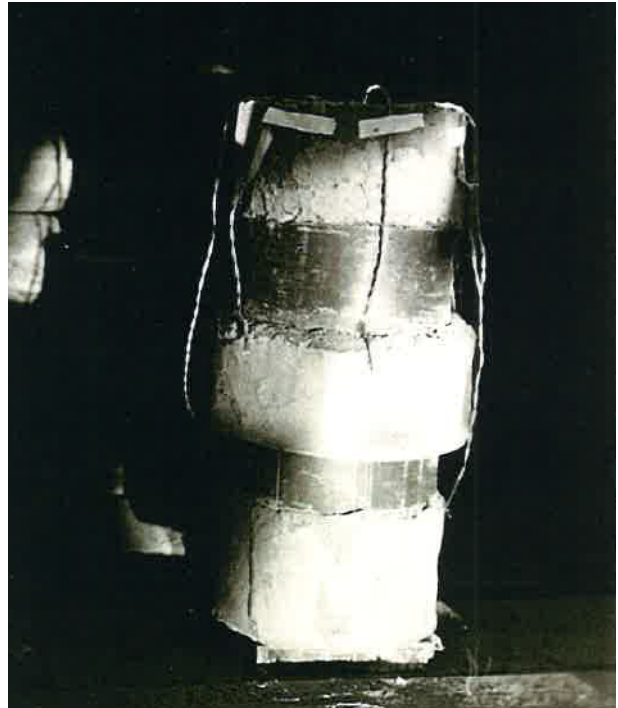

Figure 12. Concrete-reinforced elongated Jumbino ready for testing, circa late 1944.

Instead of concrete reinforcement, a second 0.6-inch thick layer of steel was added, in three segments, over the cylindrical section. The individual components were joined by a shielded metal arc weld (SMAW) technique, turned to true the circumference, reinforced with the cylindrical bands of steel, and then annealed to relieve residual stress from all the welding. Figure 13 shows an elongated Jumbino in the process of being turned down for the installation of the cylindrical reinforcing band.

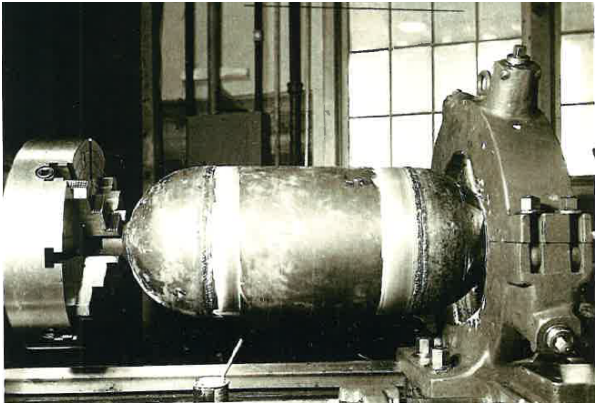

Figure 13. Fabrication of an elongated Jumbino in C shop (one of the machine shops at Los Alamos), circa late 1944. Note the small tube welded onto the end to allow fixturing on the four-jaw chuck. These were cut off after completion of the vessel.

One end of the Jumbino had a threaded access hole to load the HE. A threaded closure assembly was constructed with a feedthrough for the detonator leads, a gas relief valve, and pressure gauge, as shown in Figure 14.

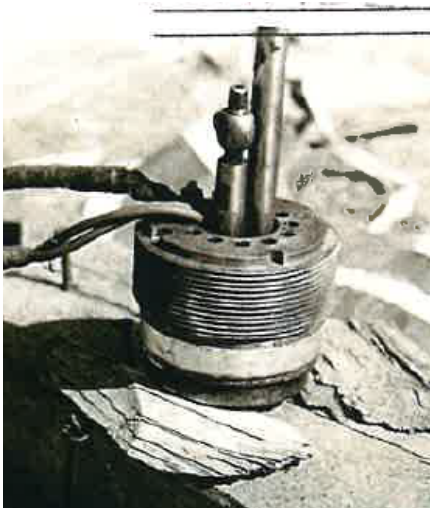

Figure 14. A Jumbino closure assembly showing lead wires, gas release valve, and bottom of pressure gauge, circa late 1944.

Remnants of elongated Jumbinos have been recovered from Two Mile Mesa, where test shots were conducted. The laminated wall construction can be seen in Figure 15.

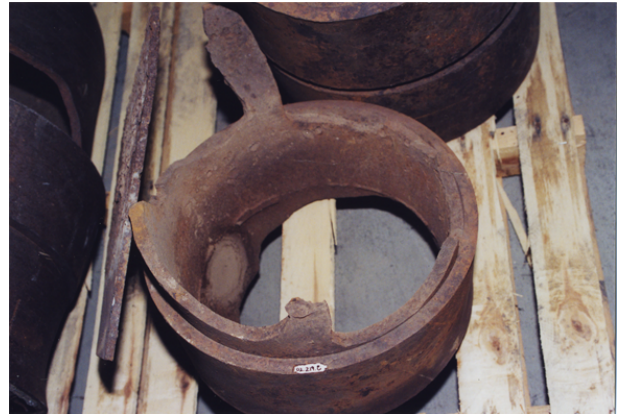

Figure 15. Laminated remnants of a Jumbino tested to failure. Picture taken in Aug 2004.

## Dumbo

A controlled hydride experiment[22] drove requirements for a larger elongated Jumbino, with charge weights up to 100 lb of HE, instead of 4.5 lb.[23] The purpose of the controlled hydride experiment is not known and, unfortunately, no additional information about the purpose was uncovered during this research project. The team nicknamed this vessel "Dumbo" and chose Babcock & Wilcox (B&W), headquarted in Barberton, OH, to fabricate it. Dumbo was constructed using alternative forming and fabrication methods as compared to the small elongated Jumbinos and the eventual Jumbo #2.[24] Instead of welding the vessel at the intersection of the cylinder and hemishell, or "head," the cylindrical section was transitioned into the head to avoid thinning about the welded in the transition region. The method had the advantage of using standard dies at the foundry versus custom fabricating a set for the project, saving time and money, time being the more important of the two. The resulting shape was not a hemisphere.

Dumbo was a 6-foot-long cylinder weighing 10 tons. It had a 4-foot ID, a 4-inch wall thickness, and an 18-inch threaded "manway"[25], which we would refer to today as a nozzle or port. A 4-1/2-inch-thick steel plate was used to spin form the heads that were finished at 3-3/4 inch to 3-7/8 inch thickness. The Project Y team did not permit B&W to cast the nozzle, so the nozzle was forged (this insistence may have been driven by the failures of the original cast-steel vessels). The individual components were welded together as shown in Figure 16. Seventy-five years later, the Dumbo vessel rests undisturbed by the elements, partially buried on the legacy IJ firing point at TA-36. Unfortunately, Dumbo has a significant amount of depleted-uranium contamination on and around the vessel, complicating recovery and inclusion with other historical artifacts of Project Y into the Manhattan Project National Historical Park because the background radiation limit exceeds those allowable for the public.

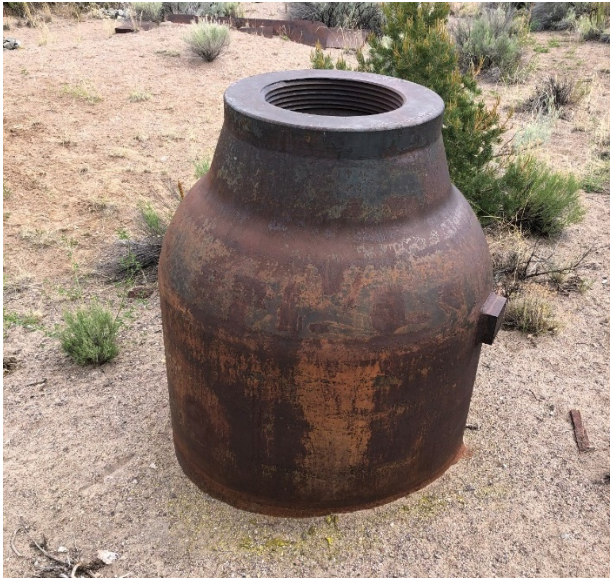

Figure 16. Dumbo resting on the IJ firing point, May 29, 2020. Note the unusual geometry of the vessel head and the uranium oxide (yellow dust) in the sand around the partially buried vessel.

## Jumbo #2: Straight Length Banded Accumulator

In keeping with the highly classified nature of the Manhattan Project, a cover story was created for the fabrication of Jumbo #2. The project devised a conceptual need for an enormous hydraulic accumulator. Thus, the official unclassified name for Jumbo #2 became "Straight Length Banded Accumulator" on drawings and procurement correspondence. This also drove obscure procurement methods so that items could not be traced back to the project. This obfuscation resulted in the Jumbo #2 procurement being handled by the War Department, United States Engineer Officer, Calexico Engineering Works, located in Los Angeles, CA, (as opposed to Jumbo #1, which never made it past a conceptual design). Based on lessons learned during the Jumbo #1 venture, R.W. Carlson came up with the basic vessel geometry and construction for Jumbo #2 in early July 1944. He called for a vessel with a 10-foot ID, 15-foot-long cylindrical section, and 10-foot-ID hemispherical heads. One end also had to include a 3-foot nozzle. These specifications became the foundation for a procurement through the War Department. By late July

1944, B&W submitted a detailed response, with drawings and a schedule for the construction of the "Straight Length Banded Accumulator" (Jumbo #2) that would meet Project Y specifications.[26]

At this point in the war, B&W was cranking out approximately 30 boilers per week to power US ships. By the end of the war, 4,100 of the 5,400 major vessels constructed for the war had B&W boilers. The most famous was the M-Type Boiler, which provided enough power to run a 25,000 hp two stage steam turbine (a high-pressure turbine combined with a low-pressure turbine). B&W's pressure vessel and boiler expertise, combined with their massive production capabilities, led to their selection as the prime contractor to build Jumbo #2 on the aggressive schedule called out by Project Y. Without even knowing it, B&W's journey into the Atomic Age began with Jumbo #2 (simply Jumbo for the remainder of this paper). In 1953, B&W established an Atomic Energy Division and provided the nuclear power components for the world's first nuclear-powered submarine, the USS Nautilus. In 1956, B&W's Atomic Energy Division opened a plant in Lynchburg, VA, for the fabrication of nuclear fuel elements, reactor cores, and other reactor parts. In 1957, B&W manufactured components for the first commercial nuclear power plant in the US, and the reactor vessel for this plant has some external similarities to Jumbo, as shown in Figure 17.

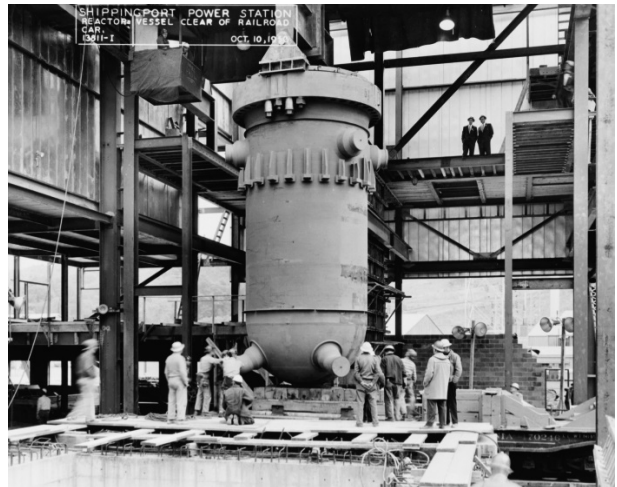

Figure 17. The Shippingport Atomic Power Station reactor vessel fabricated by B&W Atomic Energy Division, Oct 10, 1956.

B&W engineers called for Jumbo to be built from 6.5-inch plate followed by closely spaced bands shrunk in place to provide additional preloading to the structure. Not knowing the end use for the vessel gave the B&W engineers challenges, and heat effects were of primary concern. The engineers recommended full silicon-killed steel with a tensile strength of 65,000 to 70,000 psi versus 55,000 psi steels that were previously discussed with the project team. Adding silicon to the liquid steel slurry scavenges any excess oxygen, as slag, during the pour. The resulting material doesn't boil during pouring and cooling, thereby producing a more homogeneous "killed" steel. The B&W engineers highlighted that the higher-strength steel would not suffer ductility loss and that it would stand up nearly as well in a notch impact test. The higher-strength steel was proposed so the wall thickness might be reduced to 5 inches, making welding and radiographic inspection of the welds much easier. Fabrication time was quoted as four weeks for delivery of plate, 11 weeks from receipt of the first plate, and five additional weeks to apply the reinforcing rings, whether in the plant or in the field, making a total of 20 weeks for a finished Jumbo. An option for a second vessel was also part of the quote with delivery to follow the first in eight to ten weeks. Based on drawing D-67133-1 (see Appendix A for the entire drawing), the Project Y team elected to stay with a finished wall thickness of 6 inches. The exact strength of the steel is not currently known from the available historical references.

---

[26] S.L. Stewart, Memo to J.R. Oppenheimer on B&W proposal for Jumbo fabrication, 22 July 1944.

The engineers at B&W proposed a forged nozzle section, five segmented "orange peel" sections around the nozzle forging, four semi-cylindrical sections of rolled plate for the cylinder, and five additional segmented "orange peel" segments for the hemispherical head. The segments can be seen in the end view of the vessel shown in Figure 18.

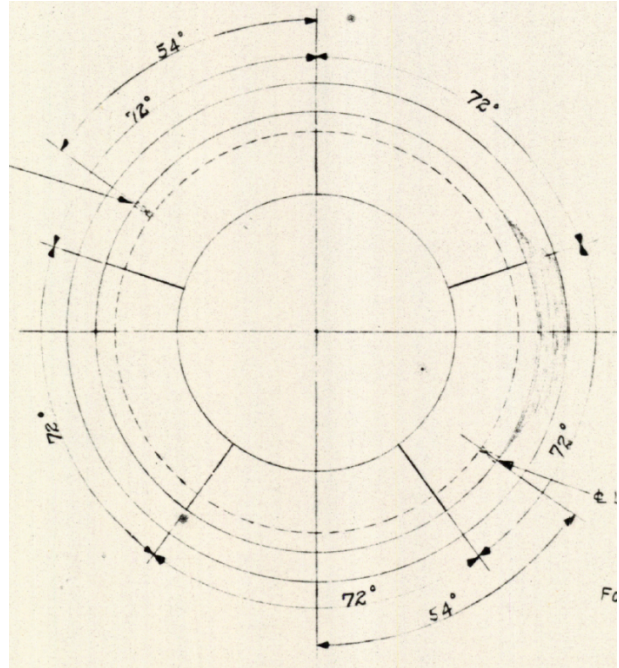

Figure 18. End view of segmented hemispheric (closed end) of Jumbo #2.

Moving from the proposal to fabrication, the entire assembly was electrically welded by a complex procedure with a combination of hand and automatic welding processes according to the American Society of Mechanical Engineers (ASME) Unfired Pressure Vessel Code. The process involved preheating the joint to 250 °F, welding the first portion (root) of the outer groove by hand, followed by automatic methods. When that was completed, the weld material that melted past the 1/8th-inch throat into the inside groove was removed (back grooving), and the welding operation was paused to magnaflux the partially completed outer groove. After passing a magnaflux test, the full 2-inch section of the inner groove was hand-welded (presumably because of a lack of access for automatic methods). At this point in the process, 3.5 inches of the 6-inch section were complete, and the joint was stress relieved at 1100 °F for three hours before the furnace was cooled to 500 °F. Once that was complete, the remaining 2.5 inches of the outer groove were automatically welded to completion. Note that the entire process was performed using cascading welding methods (skipping sections of the joint) at each step to minimize warpage. With the joint fully welded, the section was stress relieved again at 1100 °F for three hours and then furnace cooled to 500 °F. See Figure 19 for a detailed section view of the complex weld joint design necessary to weld the 6-inch-thick sections of Jumbo.

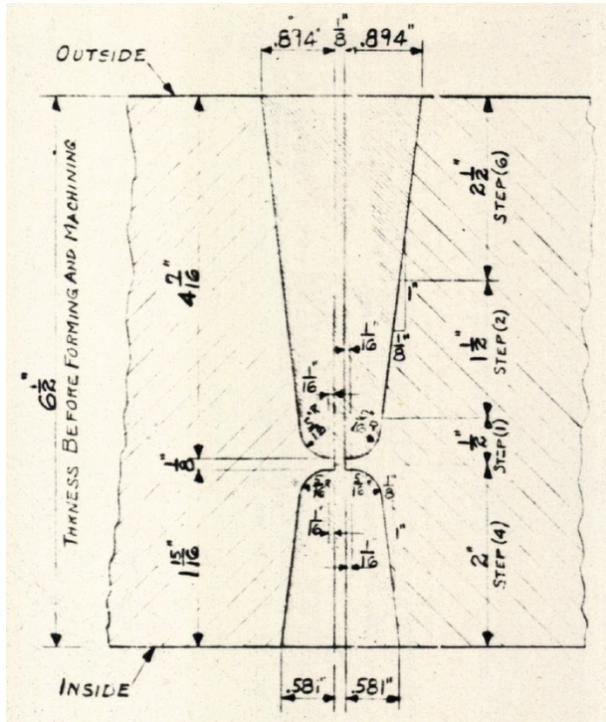

Figure 19 Section view for machining of weld joint grooves for longitudinal and girth seams of Jumbo #2. Note the 1/8-inch throat at the intersection of the inner and outer weld joints.

Once the basic vessel was completed, the outside surface of the cylinder was machined for the application of the tension banding plates. The exact thickness and construction of the banding were not determined at this point in the project, and initial requests from the Project Y team were to perform a field installation of the banding to allow enough time to determine the appropriate thickness. X-ray inspection was performed on 100% of the welds, followed by hydrostatic testing of the completed vessel at 5000 psi.

B&W recommended a hinged nozzle closure assembly for ease of use, but the Project Y team overruled and went with a robust pair of threaded plugs that locked together with 12 large studs, and corresponding nuts, in the nozzle (manway, in 1944 terms). The base diameter of the tapered nozzle was only about 29 inches in diameter. See Figure 20 for a cross section of the nozzle/plug assembly. It is worth noting that the nozzle was only large enough for the pit of the Gadget to pass through; the entire Gadget was approximately 60 inches in diameter. The entire HE assembly and case would have to be assembled inside the vessel, much like building a model ship in a bottle but on a grand scale with exposed HE charges weighing more than 100 lb each. For a historically unknown reason, almost as an afterthought, the B&W engineers recommended lining the vessel with one inch of lead. Purely speculation, but it may be possible that this was a mitigation attempt for fragments.

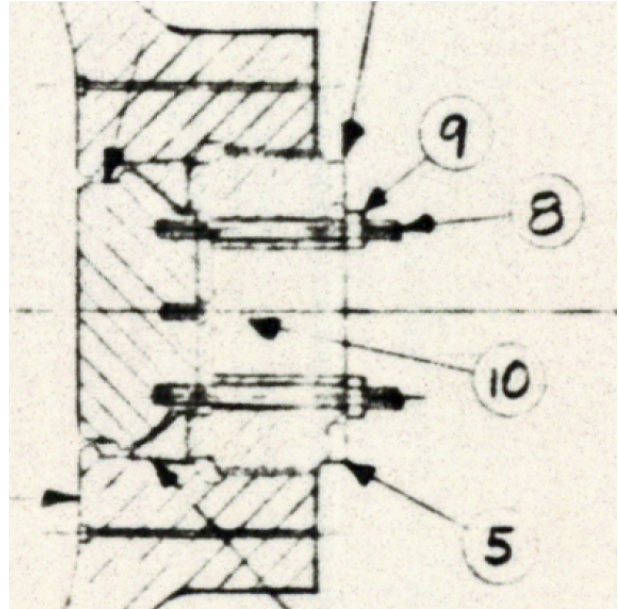

Figure 20. Section view of Jumbo #2 nozzle assembly.

Challenges were faced nearly immediately in the fabrication effort. Oppenheimer had to enlist help from General Groves to free up 200 tons of silicon-killed steel plate needed for the project because steel was in very high demand for more visible areas of the war effort, such as ships. B&W subcontracted the nozzle forging to Midvale Steel Corp, located in Philadelphia, PA, which did not succeed with the forging until the third attempt, losing considerable time in the process.[27] On a positive note, B&W adopted cautious welding techniques that produced 100% flawless welds. Given the amount of welding required for Jumbo, this was an incredible achievement and something that continues to plague vessel fabrication today. The author attributes this success to the enormous volume of work, and corresponding technical lessons learned, rolling through the B&W plant during the war. Conversely, nondestructive examination techniques and procedures have greatly improved since 1945, resulting in identification of many more flaws than would have been captured during Jumbo construction.

Also of interest was the requirement to send the partially completed vessel to another subcontractor, Mesta Machine Works, located in Pittsburgh, PA, to turn the outside of the vessel for the banding. This required moving Jumbo from Barberton, OH, to West Homestead,

---

PA,[28] where the very large lathe resided, and then back to Barberton for the banding installation and final vessel fabrication activities.

Based on teletype correspondence, it appears that B&W experienced challenges with the nondestructive weld testing on Jumbo due to the 6-inch wall thickness. In order to radiograph the welds, B&W attempted to borrow a 2-MeV x-ray machine from Army Ordnance, which was not inclined to give one up given the state of the war.[29] The B&W engineers predicted this issue, which was a primary driver for the request to use higher-strength 5-inch plate for the vessel.

As Jumbo neared completion, the banding thickness was finally determined. The team decided on 36 layers of ¼-inch-thick steel for a total of 9 inches. On top of the 6-inch base, the total thickness of the cylindrical section would be 15 inches! The bands were applied in 36-inch segments, like a stack of rings on a finger. Faint evidence of the laminated layers can be seen in Figure 21.

Jumbo underwent a final furnace stress relief before fabrication was declared complete. Following completion, hydrostatic testing was performed to 5000 psi, as specified in the requirements. Hydrostatic testing was successfully completed, and Jumbo was ready for final transportation to Pope, NM.

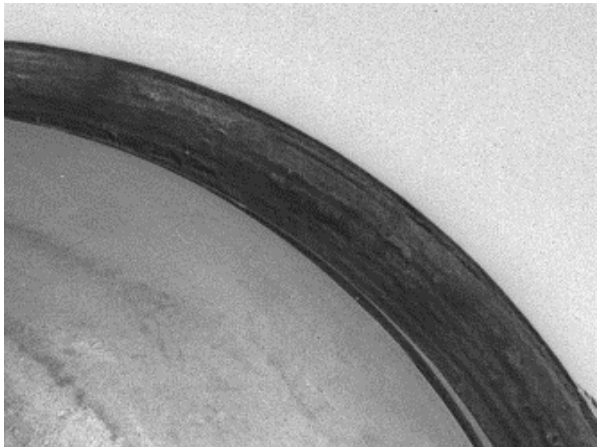

Figure 21. Thirty-six layers of laminated banding on the cylindrical section of Jumbo.

## Logistics

With the massive size of Jumbo, extensive work was done to ensure that the vessel could be transported from the fabricator to the test site. The Project Y team had already faced transportation challenges regarding maximum height and width while seeking proposals for Jumbo #1. They learned their lesson, and Jumbo #2 had a maximum diameter based on rail transportation size limits. The rail leg of the journey was solved easily by using a depressed center flat car manufactured for Carnegie Illinois Steel Corp by the Greenville Steel Car Company, located in Greenville, PA,[30] to transport ingot molds between two of their sites. The car weighed 156 tons, was 90-feet long, and had a 263-ton capacity.[31] CISX 500 is shown in Figure 22 with an ingot of steel. These types of cars were frequently used at the steel mills for heavy loads associated with steel smelting.

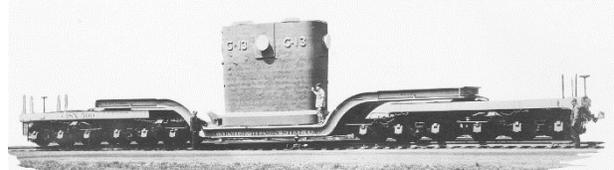

Figure 22. Car No. 500, circa 1941, was 90 feet long, 2 feet 8-3/4 inches in height from rail to the top of its depressed platform, and 6 feet 9-1/2 inches to the top of the car body; the depressed center was 18 feet long.

Various comments have been made over the years about Jumbo being the heaviest object ever moved by rail. Unfortunately, that is not possible because Southern Pacific and Union Pacific had locomotive engines that weighed far more than the 250-ton load with Jumbo. The Southern Pacific #4449 "Daylight" was built in 1941 by the Lima Locomotive Works in Lima, OH, and weighed 394 tons. Daylight is shown in Figure 23. The Union Pacific "Big Boy" engines were even bigger than Daylight. The Big Boys were built in 1941 by the American Locomotive Company in Schenectady, OH. Each Big Boy was 132-feet long, 11-feet wide, 16.5-feet tall, and weighed 675 tons! At the time of manufacture, CISX 500 was advertised as the largest freight car in the world, which may be where some of the confusion came from.

---

[28] R.W. Henderson, Status Report of Jumbo, Project Y memo, 1 January 1945.
[29] Author Unknown, B&W attempting to borrow 2-MeV X-ray for 60 days, Memo to J.R. Oppenheimer, 11 September 1944.

[30] Greenville Steel Car Company was a subsidiary of Pittsburgh Forging Company.
[31] Information from page 931 of the January 1942, Official Railway Equipment Register (ORER).

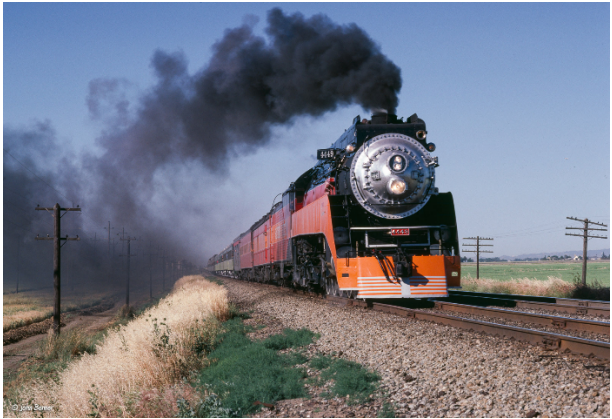

Figure 23. Southern Pacific #4449 "Daylight, circa 1981." Daylight is 110-feet long, 10-feet wide, and 16-feet tall. Daylight weighs 394 tons. Photo courtesy of John Benner.

Jumbo departed Barberton, OH, and successfully arrived in Pope, NM, by rail, as shown in Figure 24.

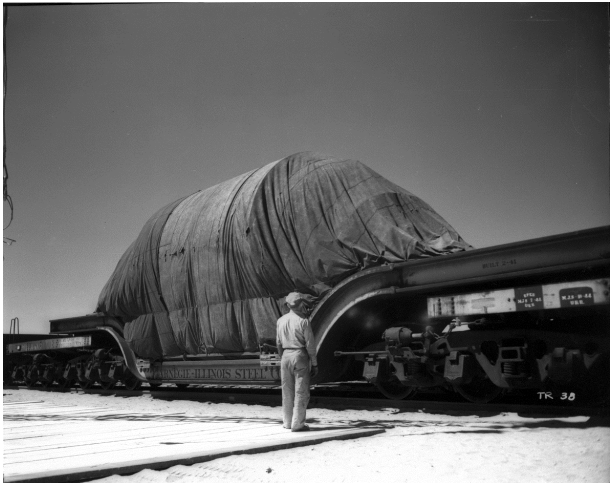

Figure 24. CISX 500 with a 214-ton "Straight Length Banded Accumulator" strapped and covered with a large tarpaulin, circa 1945.

One of the more challenging jobs was the overland journey from the rail siding to the Trinity site. In October 1944, the team met with B&W to discuss possible solutions and review project status.[32] B&W suggested a visit to John Eichleay of Eichleay Engineering Corporation, headquarted in Pittsburgh, PA. On November 14, 1944, R.W. Henderson made a visit to the Eichleay Corporation to investigate means for the overland transport leg based on this recommendation and their history of moving the penstock tube sections for the

Boulder Dam job[33] using a wheeled transporter.[34] The largest steel sections of the penstocks weighed 63 tons each, were 12-feet long, 30 feet in diameter and 2.75-inches thick, somewhat representative of the moving challenges Jumbo would present. Eichleay proposed two transportation methods, a trailer system or track system. In typical cost-saving fashion, the trailer system was cheaper at $145,000 and was the chosen solution.

The trailer method was advertised as being constructed entirely from standard components in large-scale production that would also have considerable reuse value in Army surplus stock.[35] The trailer was designed so that it could be broken down into several units for transportation to the site.[36] The "standard components in large scale production" Eichleay referenced in the quote were manufactured by the Rogers Brothers Corporation, based in Albion, PA.[37] Rogers Brothers sold their first trailer in 1914, and they are one of the few companies that supported the Manhattan Project that is still in business today. The Army issued a contract for the Jumbo trailer with a delivery deadline one month later! Fortunately, the company was already building tank retriever trailers with a similar axle configuration and was busy churning out ten trailers a day. Rogers finished the trailer on time, and the unassembled trailer was delivered to Pope by rail. Figure 25 shows the trailer after the move was completed.

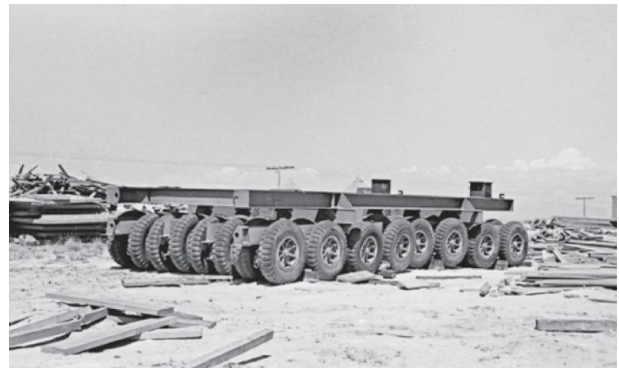

Figure 25. The Rogers-built trailer for Jumbo, 18-feet wide by 40-feet long with 64 desert-type tires. Photo is post Trinity, circa late 1945.

The trailer was assembled next to the Pope railroad siding, and with the help of three Caterpillar D-8 tractors, it successfully hauled Jumbo to the Trinity site, as shown in Figure 26. Roger Brothers Corporation still manufactures trailers, and the Jumbo trailer is still the largest trailer ever built by the company.

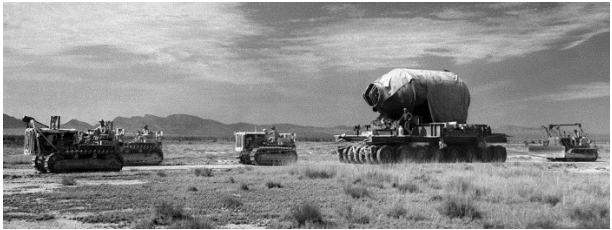

Figure 26. Three Caterpillar D-8 tractors pull the Rogers trailer to the Trinity site, while a single Caterpillar bulldozer follows as the brakes for the massive load, circa mid 1945.

Figure 27 is a map of the firing site and identifies the route across the desert to the resting place for Jumbo. The Pope siding is along the left side of the figure, and the final resting place of Jumbo is near the center.

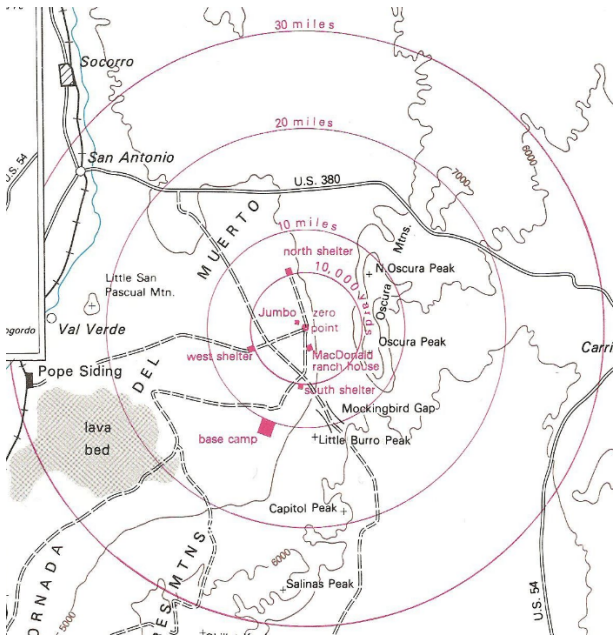

Figure 27. Location of Jumbo 45 degrees north-west of Trinity zero point. Note the location of Pope siding and the 30-mile route Jumbo had to travel across the desert.

## Trinity

Major concerns on the viability of the Gadget were reinforced when a full-scale hydrodynamic test was performed at TA-18 three days prior to the planned execution of Trinity, and initial analysis of the data indicated the Gadget would not work. Despite the lingering concerns, the decision to not use Jumbo had already been made over concerns of optical access for high-speed motion pictures, continuous spectrographic studies, and observations of the shock wave.[38] This data was critical for the scientists to understand the function of the Gadget. Calculations were made on the energy necessary to vaporize Jumbo and all of the contents, but these calculations proved unnecessary. By March 1945, Jumbo and the erection tower (somewhat representative of a multistory building) were relegated to become part of the world's first nuclear effects test, versus the intended purpose of containing the precious plutonium in the event of a fizzle.[39]

Jumbo was transferred from the Rogers trailer to cribbing at the base of the tower, and a block and tackle was used to hoist it from horizontal to vertical inside the tower. Emplacement of Jumbo was the sole purpose of the tower. Figure 28 shows Jumbo being prepared for the foundation pit. Jumbo was lowered in the tower until the laminated reinforcing bands were level with the top of the foundation forms (approximately level with the surrounding terrain).

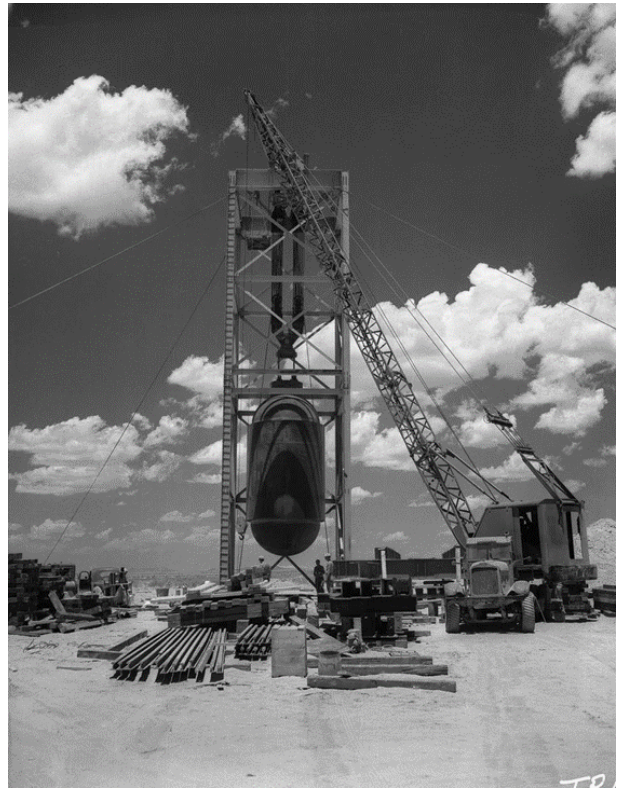

Figure 28. Jumbo suspended by a large block and tackle in the tower, circa summer 1945. Note the missing x-bracing where Jumbo was hoisted from horizontal at the base of the tower to vertical within the tower.

After lowering, concrete was poured around the base of Jumbo, providing a solid foundation for the test. Once the concrete cured, all the rigging was removed, and Jumbo was ready to take a beating unlike anything the world had ever seen. Figure 29 shows Jumbo in the preshot configuration.

---

[38] J.E. Mack, "Disadvantages of Jumbo for Optical Tests," Project Y Memo, 28 February 1945.

[39] K.T. Bainbridge, "Construction at Trinity," Project Y Memo, March 15, 1945.

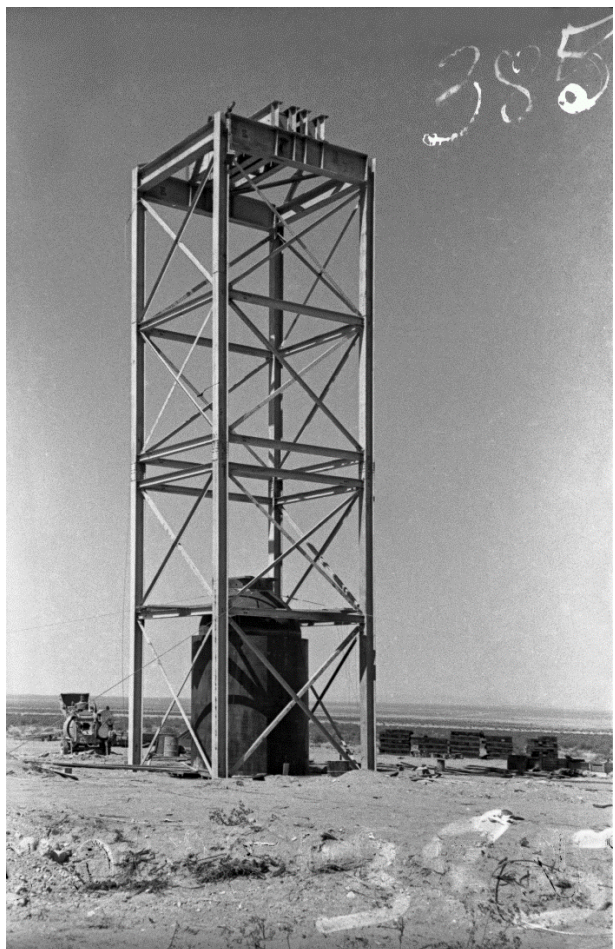

Figure 29. Jumbo in preshot configuration, circa summer 1945.

The Gadget was fired on July 16, 1945, at 0529 from a 100-foot-tall tower 800 yards away from Jumbo and generated a 25-kiloton (+/-2 kt) yield[40]. Based on evidence of the preshot wagers, the actual yield was far greater than most of the Project Y team predicted, including Oppenheimer.

The Gadget shot tower was vaporized in the fireball, the desert sand was melted into green glass approximately 400 yards in all directions, the heavy-duty tower used to emplace Jumbo was blown down, and yet Jumbo emerged unscathed—a testament to the strength and durability of the vessel. Figure 30 shows the twisted wreckage of the tower with Jumbo front and center.

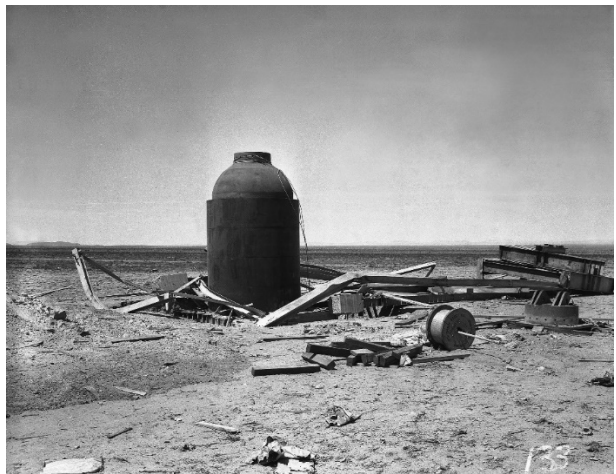

Figure 30. Postshot survey of Jumbo, July, 1945. Note the complete destruction of the heavy-duty erection tower.

## Demolition Operation at Trinity Site

Jumbo's tower was completely mangled by Trinity but Jumbo survived unscathed and remained vertical on the concrete foundation.[41] For the Project Y team, this was a major triumph, and the team speculated about future uses of Jumbo.

With this in mind, the team had a watertight cover installed over the nozzle and a tarpaulin placed on top of the nozzle cover, secured with steel banding tape, to protect against corrosion damage. The remaining vessel components were cleaned, greased, stacked, covered with a tarpaulin, and labeled. The weather-sealed Jumbo can be seen in Figure 31.

---

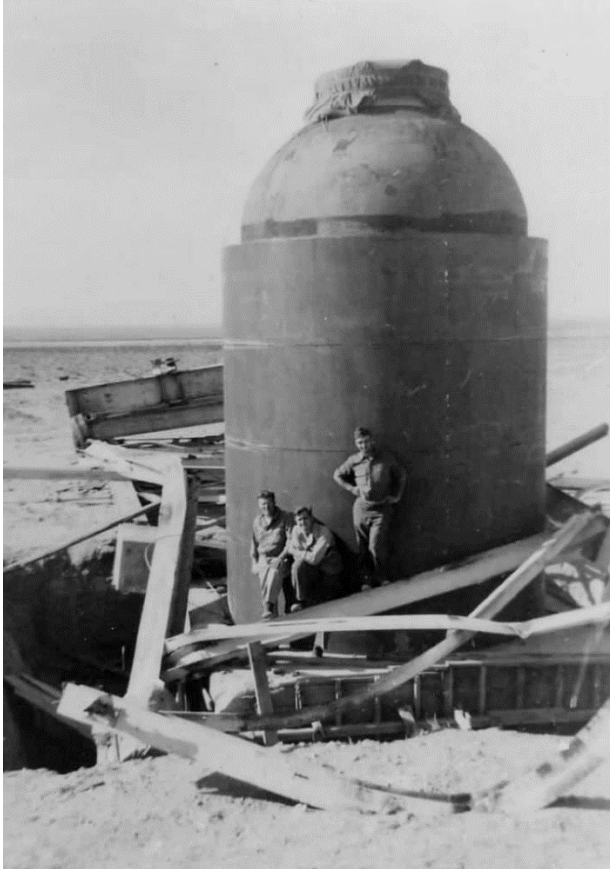

After a few months of peace and quiet in the Jornada Del Muerto desert, Army personnel assigned to Trinity decided that Jumbo would be a convenient place to destroy eight unserviceable AN-M64 500-lb General Purpose Bombs, as shown in Figure 32. The AN-M64 contained 267 lb of TNT (or a similar quantity of Amatol or Comp B). The HE was contained within a 0.3-inch-thick steel case that was 18 inches in diameter and 57-inches long. Nothing compared to the explosive force in the Gadget, but it was still a formidable challenge.

The bombs were lowered into the bottom of Jumbo, and the finished assembly contained a combined HE load of 2136-lb TNT plus an unidentified weight of Composition C-II plastic explosive. C-II was placed in the fuse cavity of six bombs with one #6 electric blasting cap each, and additional C-II was wedged between bombs, fused and unfused, to ensure high-order detonation.

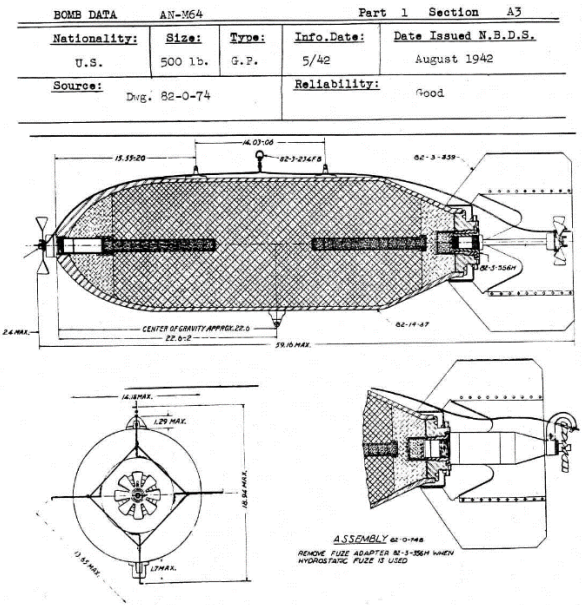

Figure 32. The AN-M64 500-lb General Purpose Bomb, August 1942. Note forward and aft fuze locations. Composition C-II was packed in the unfused bombs.

The Army decided to proceed with the little experiment without consulting the Project Y engineers who designed Jumbo. The complex nozzle plug was not installed in Jumbo, i.e., it was left open to vent. Based on the robust construction of Jumbo, the firing site was set up a mere 600 ft from the charge, behind a concrete wall on an old instrument vault. The officer in charge was told that fragmentation was not expected because Jumbo had been constructed to retain a much greater charge weight. All personnel and equipment, except a two-person firing party, were cleared to a one-mile radius. The charge was fired at 1130 on April 16, 1946. The result is not what the team expected. The foundation was pulverized and scattered over a large area. Both ends were torn off Jumbo, and fragments were thrown as far as three-quarters of one mile. The cylindrical section landed on its side in a crater that concealed all but a few feet of the twenty-foot-tall vessel when viewed from a distance.[42] Recently declassified photos of the aftermath are shown in Figures 33–35. Much speculation is still made about the defacement of Jumbo, but the evidence is clear. The team did not intend to completely destroy the vessel, or else they would not have located the firing location a mere 600 feet from the vessel. The two-person firing party was lucky to escape the incident alive, and the event certainly constituted a near miss! Having read many of the speculations regarding the defacement of Jumbo, the author has a slightly different explanation. From the other

---

speculations, it appears that a justifiable use for Jumbo was strongly desired in case auditors dug into the project. A vessel for the safe destruction of unserviceable bombs or Un-Exploded Ordnance (UXO), like modern bomb disposal trailers, would have provided that justification. Trinity was conducted on a bombing range, with other near misses, so UXO was a very real problem at the range, and remains so today.

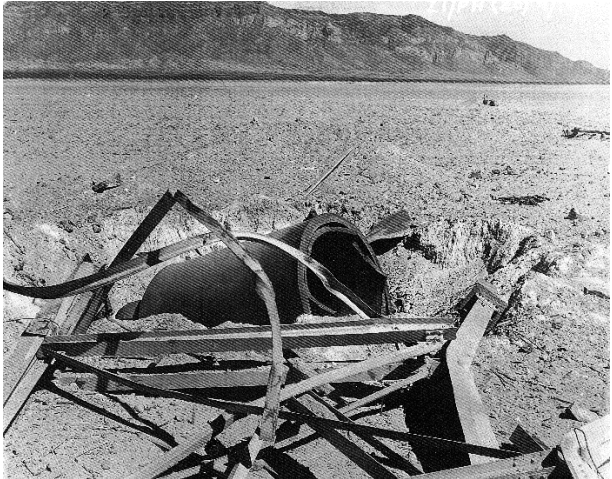

Figure 33. Aftermath of the Army experiment in Jumbo, April 16, 1946.

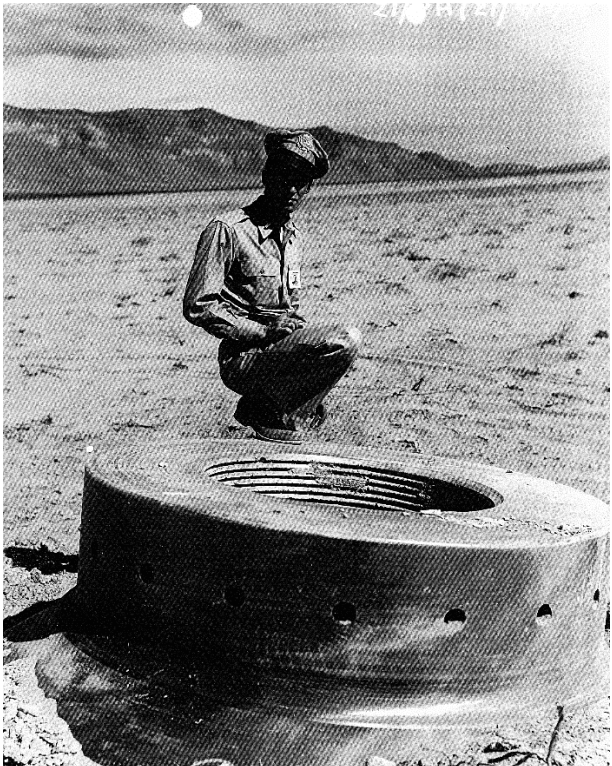

Figure 34. The Jumbo nozzle forging laying on the desert floor after the Army experiment, April 16, 1946. Note the diameter and threading of the nozzle.

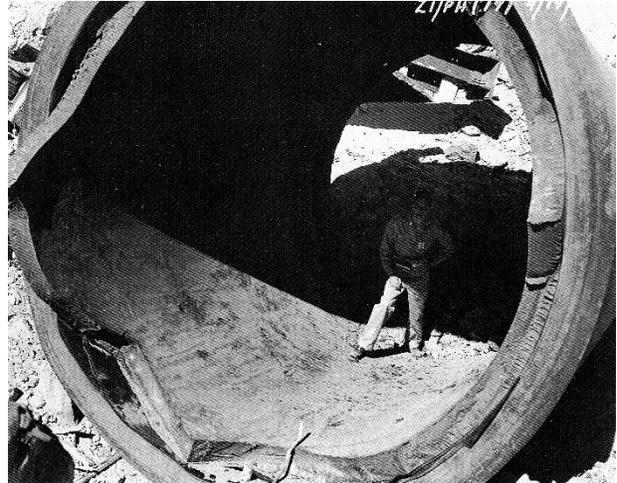

Figure 35. View of the fractured body of Jumbo, April 16, 1946. Note that the vessel did not fail at the welds but in the parent material, providing clear evidence of the welding process quality.

R.W. Henderson provided a rather terse letter to Lt. Col. Frolich admonishing them for the configuration of the "test." Henderson stated, "It is of course realized that the manner in which the bombs were fired in the vessel completely nullified the design basis but there are certain aspects of the manner of failure which are of considerable interest. During the course of our experiments we fired reduced charges in off-center geometry in the vessel which resulted in much the same type of failure even though fully scaled charges located dead center were retained by the vessel."[43]

### Jumbo #3

With the establishment of Los Alamos Scientific Laboratory (LASL) in 1947,[44] GMX Division was interested in a closed chamber to conduct HE experiments.[45] The concrete chamber at Q-Site was not performing as desired at 3 lb of explosives, and GMX-8 desired a chamber for the purpose of indefinite firing up to 10-lb of HE with protection from high wind, rain, and sunlight.[46] The concrete tube at Q-Site failed after 50 shots at an average charge of 3-lb TNT. The pit at L-Site had proven to handle up to 30-lb of HE in one charge, but the earth fill around the chamber had to be reworked after five shots. Protection from the sun was desired for spectrographic work and flash photography. Rain protection was desired to ensure electrical pins operated correctly. Wind protection was desired to maintain shot alignment, and general adverse weather protection was desired so that scheduled shots were not dependent on weather conditions. The group had $150,000 to come up with a solution.

The first option was to refurbish Jumbo #2 with slight modifications. They wanted a manway on each end of the vessel for better access. The group went back to Trinity with B&W in September 1949 to see if it was feasible to repair the cylindrical section of Jumbo, attach new heads, and develop an estimate for the task. B&W engineers evaluated Jumbo and determined that it was possible to repair Jumbo. They provided a quote to GMX for $101,300 (with all options) to complete the refurbishment and transport from the Trinity site in New Mexico to Ohio and then back to Santa Fe.[47]

In addition to the refurbishment approach, GMX-8 obtained quotes for the fabrication of a brand new vessel from several companies, including B&W. The B&W quote for Jumbo #3 (historical documents simply refer to Jumbo, but to avoid confusion, this vessel is identified as Jumbo #3) was $165,500.[48] B&W Drawing No. HD-75681-0 for the Jumbo #3 quote is attached in Appendix B for reference. A.O. Smith, headquartered in Milwaukee, WI, won the bid and began fabrication of Jumbo #3. Just as B&W experienced problems during the fabrication of Jumbo #2, A.O. Smith experienced some

challenges and accompanying schedule delays. Steel plate received from the mill had laminations in the plate from the rolling and shearing process used to form the metal, but this was not known upon receipt of the material. In a similar fashion to the fabrication of Jumbo #2 by B&W, A.O. Smith utilized Submerged Arc Welding (SAW)[49] for the vessel sections followed by x-ray inspection. During the x-ray inspection of one vessel head, cracks were found in the seams. The steel plate laminations provided stress concentrations and under thermal stress from the welding, they cracked. The cracks ranged from 7-inches to 29-inches long and they were up to 1-1/2 inches deep. All lamination defects and resulting defective welds had to be removed and welded again with much slower manual welding procedures. Some repairs requiring four attempts before succeeding.[50] Fortunately, the cylindrical section was fabricated successfully on the first attempt. It is worth noting that the vessel had 3,456-linear inches of weld. The exact number of welding passed required for each 6-inch thick joint is not known. Based on information available for 2-inch thick sections, the author estimates 200 welding passes for each joint/seam. With this assumption, the total length of weld would be around 691,200 linear inches or nearly 11 miles!

Jumbo #3 was shipped by rail from Milwaukee, WI, to Santa Fe, NM, without incident and handed off to Lowdermilk Bros. Construction Company, located in Espanola, NM, for the road journey from Santa Fe to Los Alamos. During the move of Jumbo #3 to Los Alamos, Lowdermilk welded reinforcements at 22 different locations on the vessel to brace it in the transportation structure. The severely overloaded trailer can be seen in Figure 336. These welds created a severe crack in the end of one forging and several minor cracks at the edge of several other welds. Since all of the welding was done on the outside of the vessel, Lowdermilk believed that the stresses that were created from the welding would be minor. Unfortunately, this was not a correct assumption.

---

[43] R.W. Henderson, "Destruction of Jumbo – Trinity," Z-4 Inter-Office Memo, May 10, 1946.
[44] In 1981, Los Alamos Scientific Laboratory was renamed Los Alamos National Laboratory.
[45] A.W. Campbell, "JUMBO," LASL Memo, 10 October 1949.
[46] A.W. Campbell, "Utility of a Jumbo-like Chamber," LASL Memo, 9 February 1950.
[47] C.H. Gay, B&W quote to repair the large pressure vessel at the Trinity Area, October 11, 1949

[48] R.C. Ruth, "Special Heavy Banded Vessel," B&W Memo, December 20, 1949.
[49] Submerged Arc Welding (SAW) was developed by National Tube Co. in McKeesport, PA by Boris S. Robinoff in 1929 and patented in 1930. The rights were sold to Linde Air Products and trademarked as UNION-MELT.
[50] M.W. Breck, University of Calilfornia MV-1439, A.O. Smith Corporation Memo, June 6, 1951.

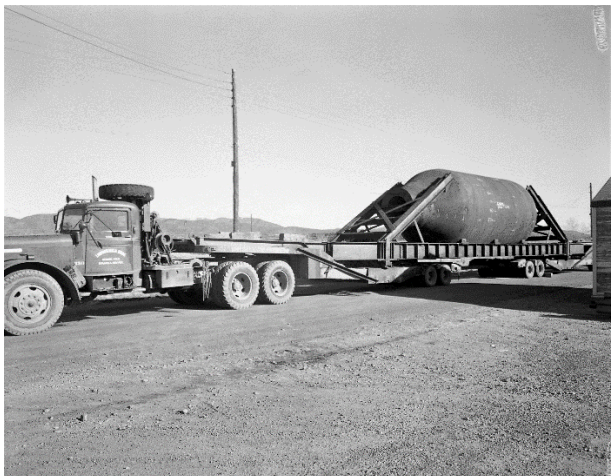

Figure 336. Lowdermilk Bros. Construction Company's overloaded trailer in Santa Fe, NM on January 24, 1952. Note the additional structure welded to Jumbo #3 to stabilize the load.

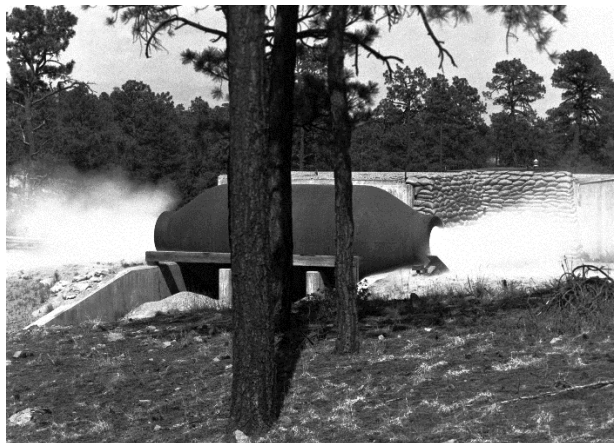

Figure 347. Jumbo #3 in service at Eenie Site, circa 1953. Note the elongated heads, dual manways, and concrete support cradles.

A.O. Smith was called in to inspect the vessel. They stated, "The reason for the severe crack in the end forging was readily apparent, because of the higher carbon and manganese content of the end forging, which was AISI 1330 specification. The remainder of the vessel was fabricated from ASTM A-212 steel specification. The welding performed by the contractor was done with an all-position type of welding electrode, without resorting to preheat prior to welding."[51] A.O. Smith recommended cutting the structure off with a gas torch, grinding the welds flush to the surface of the surrounding steel then carefully burnishing and polishing the areas for inspection. Inspection included magnetic particle and visual inspection using etching and a magnifying glass. They recommended that any defects found be removed by additional grinding. If grinding was required deeper than 1/4-inch deep then weld repair would be necessary. The welding procedure entailed preheating the areas uniformly to 350 °F, welding with 3/16-inch diameter SW-64 electrodes, and peening each weld bead up to the last layer. The final layer was not to be peened. After completion of welding, the excess material would be removed by grinding until it was flush with the surrounding surface. Following this procedure, the areas were magnafluxed and etched again to determine the soundness of the repair. Fortunately, upon removal of all the moving structure and inspection of the welds, no additional defects were found.[52]

Jumbo #3 was successfully repaired and put into service at the Eenie firing site, as shown in Figure 347.

After approximately two decades of use, LASL personnel decided that the Pulsed High Energy Radiographic Machine Emitting X-Rays (PHERMEX) needed a camera protection structure more than the Eenie firing site needed a semi-enclosed firing chamber. Waterman Incorporated, a local Los Alamos moving contractor, was hired by LASL to move Jumbo from Eenie site to PHERMEX. To do this, Waterman created a steep dirt road, with a 14% grade, from Eenie up the mesa to PHERMEX. The road was appropriately named Jumbo Road. Just like the logistical challenges with the original Jumbo, Waterman faced challenges with the project when wood cribbing gave way, and Jumbo #3 almost fell off the trailer. See Figure 358 as Jumbo #3 was moved to PHERMEX.

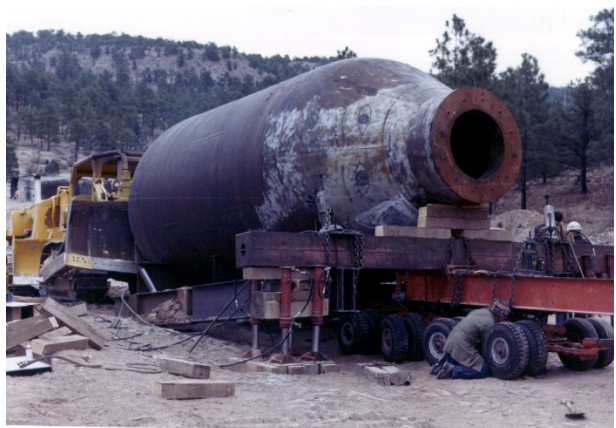

Figure 358. Jumbo #3 being moved from Eenie to PHERMEX by Waterman Incorporated, circa 1976.

Jumbo #3 was successfully moved to PHERMEX but never served as camera protection because Fred Dormire, the prime investigator, passed away before the camera was built. Figure 39 shows Jumbo #3 with a new set of legs at its intended camera-protection location at PHERMEX. This location got in the way of other experiments, so the vessel was "crab walked" off the firing point using the site crane. Jumbo #3 had to be moved in this fashion because it was too heavy for the crane to lift completely off the ground.

In 1992, LANL Group M-4 considered using Jumbo #3 to protect the Ector radiographic machine from TA-15-306, which would be situated perpendicular to the PHERMEX beam axis. The project was referred to as the "Relocate Ector to PHERMEX" project. The intent of the project was to prove the viability and benefit of dual-axis radiography to help motivate the research and development of a state-of-the-art dual-axis linear accelerator (now known as DARHT). Ultimately, the decision was made to pursue a different protective envelope for the "Relocate Ector to PHERMEX" project and Jumbo #3 stayed at PHERMEX. For a number of reasons not relevant to this paper, the project was cancelled in 1993.

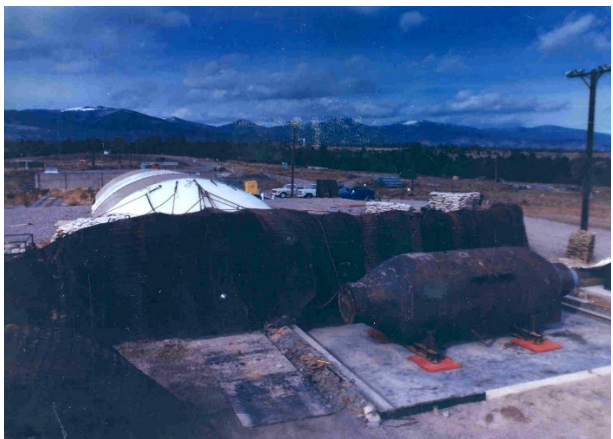

Figure 369. Picture of Jumbo #3 at the PHERMEX firing point, circa 1977.

In 2008, DARHT personnel evaluated moving Jumbo #3 from PHERMEX to DARHT so that much larger shots could be fired on the DARHT firing point, but the move never happened.[53]

Finally, in 2014, LANL Group WX-3 investigated putting Jumbo #3 back into service at LANL's R306 firing site. The intent was to provide protection from the weather and protect the surrounding forest from hot fragments that some shots create. The investigation stopped when the expense of moving and retrofitting it for current diagnostics could not be justified compared to the cost and relative ease of using modern 6-foot or 8-foot-diameter confinement vessels. Only time will tell if Jumbo #3 has another life. Meanwhile, Jumbo #3 will wait patiently on the edge of the PHERMEX firing point for another adventure or the occasional visit from curious site personnel, as shown in Figure 40.

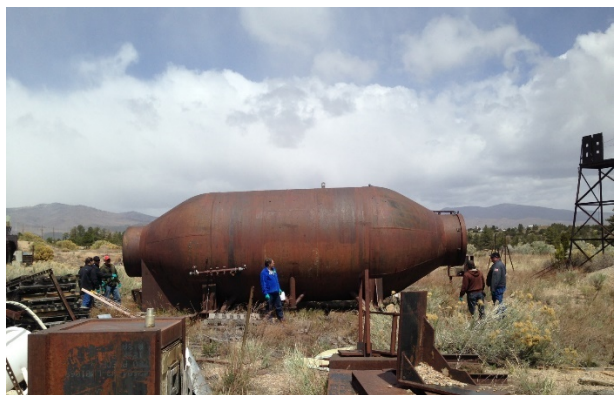

Figure 40. Jumbo #3 resting on the side of the PHERMEX firing point, May 1, 2014. Note the field-welded legs that may have created new damage to the vessel. History repeats itself.

**Essential Tools**

The Jumbo saga—Jumbos #1, #2, and #3—opened the door to an entire regime of testing methods that became critical for nuclear weapon development in the Cold War and today's science-based stockpile stewardship. The steel vessel project led by Project Y personnel was the first known investigation into near-field blast pressures generated by large charges and the first attempt to confine an HE detonation. Jumbo was the first object subjected to a nuclear detonation (see Figure 30, after the Trinity test) and emerge from the event unscathed. Because this was the world's first nuclear detonation and because of the damage done to Jumbo by the Army, Jumbo is now a popular artifact at the Trinity site, and the two Jumbos remain as the largest "portable" vessels manufactured by the US to confine explosive detonations. As big and impressive as the Jumbos are, the Russian Asgir sphere, shown in Figure 41, is in another league.[54] The sphere was originally designed to contain large conventional-explosive events for the Arzamas-16 nuclear weapons laboratory and was shown to LASL Director Harold Agnew during his USSR visit in 1975.

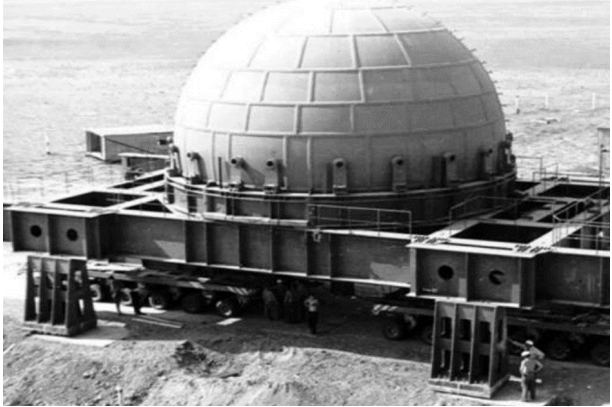

Figure 41. The Russian Asgir sphere, circa 1975. Note the two men in the lower right corner for scale.

A review of these amazing engineering accomplishments highlights the broad spectrum of challenges that the team overcame to design, fabricate, move, and use confinement vessels. Today our engineers face many of the same challenges as confinement vessels are designed, fabricated, and used to support essential national security efforts at Los Alamos National Laboratory, Lawrence Livermore National Laboratory and the Nevada National Security Site.

This work was supported by the US Department of Energy through the Los Alamos National Laboratory. Los Alamos National Laboratory is operated by Triad National Security, LLC, for the National Nuclear Security Administration of the US Department of Energy under Contract No. 89233218CNA000001.

**Appendix A – B&W Straight Length Banded Accumulator**

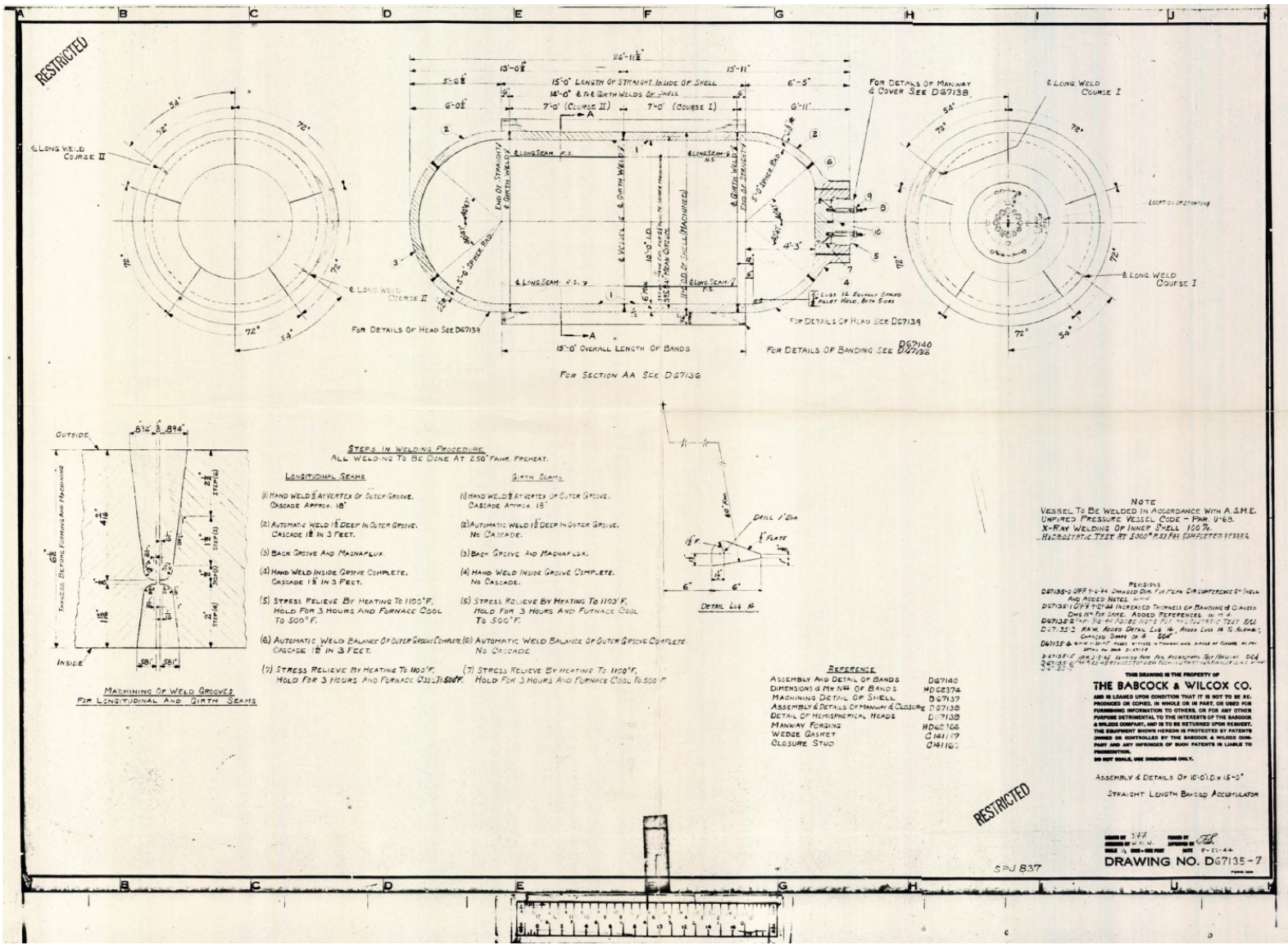

**Appendix B – B&W Proposed Pressure Vessel for Jumbo #3**

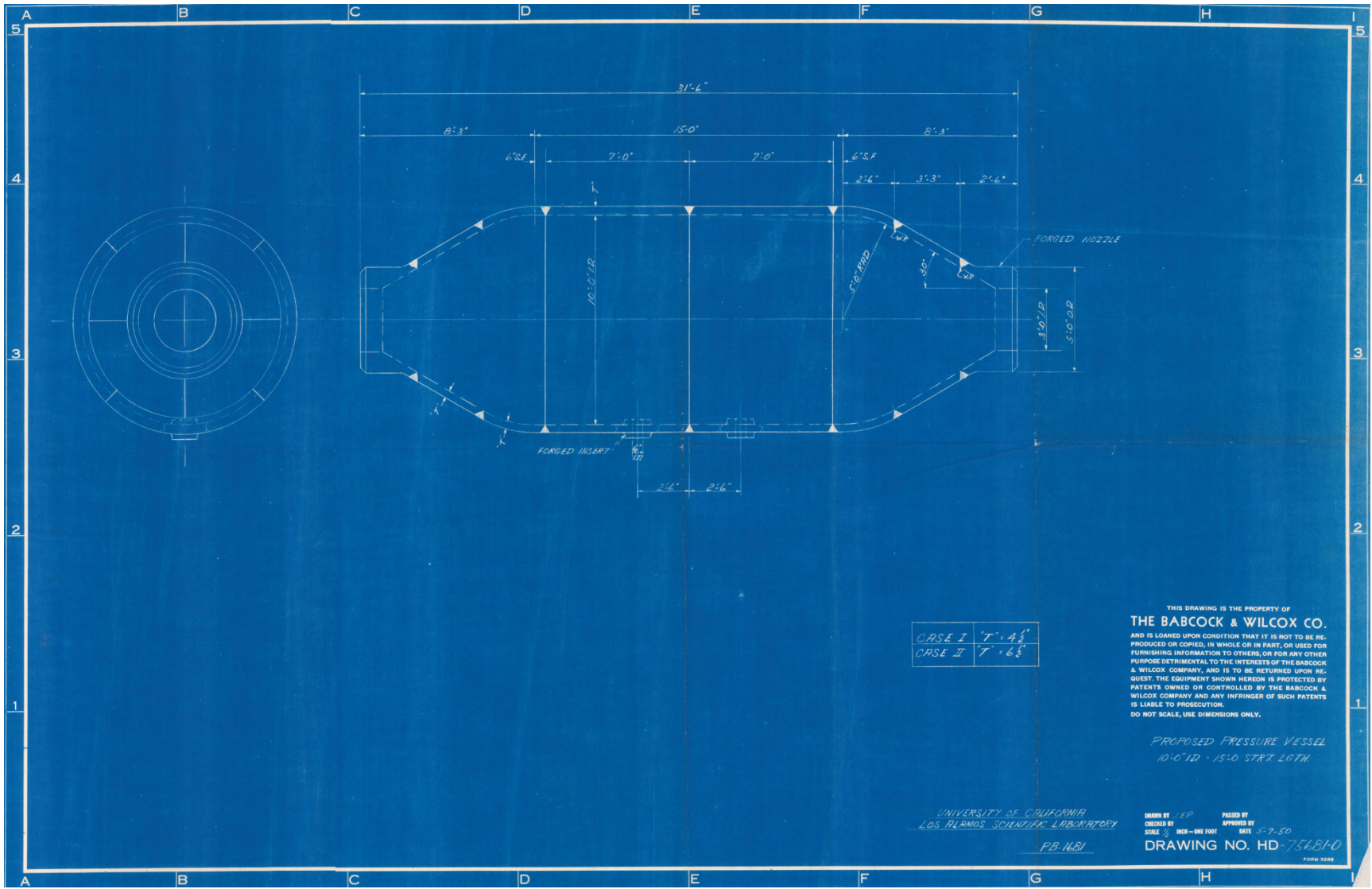